\documentclass[preprint,12pt]{elsarticle}
\usepackage{amssymb}
\usepackage{amsmath}
\usepackage{url}
\usepackage{hyperref}
\usepackage{subcaption}
\journal{Physics of the Dark Universe}

\begin{document}

\begin{frontmatter}
    \title{\boldmath{Evolutionary optimization of cosmological parameters using metropolis acceptance criterion}}
    \author[1]{Supin P Surendran\corref{cor1}}
    \ead{supinps@gmail.com}
    \author[1]{Aiswarya A,}
    \author[2]{Rinsy Thomas}
    \author[3]{and Minu Joy}
    \cortext[cor1]{Corresponding author.}

    \affiliation[1]{
        organization={School of Pure and Applied Physics, Mahatma Gandhi University},
        addressline={Priyadarshini Hills, Kottayam},
        postcode={686 560},
        state={Kerala},
        country={India}
    }
    \affiliation[2]{
        organization={Department of Physics, CMS College},
        addressline={Kottayam},
        postcode={686 001},
        state={Kottayam},
        country={India}
    }
    \affiliation[3]{
        organization={Department of Physics, Alphonsa College},
        addressline={Pala},
        postcode={686 574},
        state={Kottayam},
        country={India}
    }

    \begin{abstract}
        A novel evolutionary method is introduced that can be used for constraining the parameters and theoretical models of Cosmology. The newly proposed algorithm, which is inherently parallel by design, is able to obtain the full potential of multi-core machines. With this algorithm, we could obtain the best-fit parameters of the $\Lambda CDM$ cosmological model as well as the uncertainties and identify the discrepancy in the Hubble parameter $H_0$. In the present work we discuss the design principle of this novel approach and also the results from the analysis of Pantheon, OHD and Planck datasets are reported here. The estimation of parameters shows significant consistency with the previously reported values as well as a higher computational performance measured in number iterations compared to the other similar exercises.
    \end{abstract}

\end{frontmatter}


\section{Introduction}\label{section:intro}
One of the most important exercises in observational cosmology is to infer the cosmological parameters from the data acquired through various projects like Wilkinson Microwave Anisotropy Probe (WMAP)\cite{Hinshaw2013},   Planck\cite{Tauber2004}, BOOMERANG\cite{crill2003boomerang}, MAXIMA\cite{rabii2006maxima}, ACT\cite{aiola2020atacama}, SPT\cite{chown2018maps} and Sloan Digital Sky Survey (SDSS)\cite{York2000} and many other. The data thus obtained reveals the dynamics of the universe and its evolution only within the constraints of uncertainties. Thus myriads of models, which explains the physics of the universe, were proposed throughout history. A cosmological data analysis program is tasked with the estimation of cosmological parameters against such models to discriminate and constrain different cosmological models. Bayesian analysis and its implementation -- Monte Carlo Markov Chain (MCMC)\cite{brooks2011handbook}, is a widely used parameter estimation technique in general. Among them, {\textit{CosmoMC}\cite{Lewis_2002}}, \textit{MontePython}\cite{Audren:2012wb}, \textit{CosmoSIS}\cite{zuntz2015cosmosis}, \textit{Cobaya}\cite{torrado2021cobaya} are successful  MCMC frameworks in cosmological data analysis. The robustness of these Metropolis-Hasting (MH)\cite{metropolis} criterion-based codes are apparent in most of the literature in observational cosmology. The Cosmic Microwave Background radiation (CMB) data analysis mainly deals with the interpretation of CMB data and its impact on the understanding of the inflationary era, whereas Supernovae (SNe) data is about the evolution of the present universe. Both the paradigm aim to fix the classical cosmological parameters within the error bar as dictated by the fundamental notions of physical cosmology. However, the former when compared with SNe analysis comprises a much larger parameter space, making the problems under scrutiny more problematic.

Sampling techniques that implement MCMC sampler, apart from their wide popularity, carry some critical problems within the algorithm as pointed out by Geyer\cite{geyer_practical_1992} and Raftery\&Lewis\cite{raftery_practical_1992}. A few to mention are the ambiguity in the determination of convergence, especially when the parameters are having a significant amount of correlation, possible pollution from the correlation within the chain and the computation time required to reach a stationary equilibrium of the chains. MCMC techniques are hard-to-parallelize algorithms by design since the inherent iterative nature requires each iteration in the chain to depend on the previous iteration. However, some efforts have been made to parallelize the algorithm to reduce the computational overhead. Another key drawback is in the sampling process which scales to O(n) computation for an n-dimensional parameter space in each step of the chain.

In the present work, we propose a novel computation approach to estimate cosmological parameters. The evolutionary algorithm we discuss here, adopts the advantages of MCMC sampling technique without losing the computational performance of the optimization methods. Having the design principle laid upon that of evolutionary optimization\cite{simon2013evolutionary}  technique, it lets the algorithm to have a much larger search space and faster convergence. Also the evolutionary algorithm does the optimization with a large population of parameter candidates but with limited amount of sampling. Here we estimate the parameters for $\Lambda CDM$ model of cosmology corresponding to both present and early stages of universe with this algorithm. The parallel evolution and optimization of parameters allows the algorithm to confront the problems due to the  dimensionality of the parameter space without much computational difficulty. Thus eliminating the need for strict constraints on the parameters in the prior assumption and in the further sampling.

Through this work we attempt to describe the possibilities of a hybrid methodology for analyzing the cosmological dataset and its advantages in estimating cosmological parameters. Here, we consider $\Lambda CDM$ cosmological models which has much cosmological importance in identifying the discrepancies in various observations and theoretical considerations. In particular, the work is focused on finding out departures in parameter values when different sets of observational data are being used against $\Lambda CDM$. The paper is divided as follows. The Section \ref{section:theoryobservation} provides background information on the fundamental notion of theoretical framework of the Cosmology, particularly $\Lambda CDM$. Various datasets used in this work are also decribed in the same section. In Section \ref{section:datsets}, we discuss about various available cosmological data set and compilation projects in detail. The section also explains the nature of the observables and different uncertainties associated with the data.  A formal description of the MCMC framework and its applicability is portrayed along with algorithmic description in Section \ref{section:methodology}. In  the same Section the design and application of evolutionary optimization technique in detail is also described. The results of our method is presented in the Section \ref{section:results} along with plots describing the process of parameter estimation. In Section \ref{section:conclusion}, we conclude with a discussion on the strength and weaknesses of the evolutionary optimization algorithm.

\section{Cosmology, Theory and Observations}\label{section:theoryobservation}
A cosmological model describes the physics of the universe and its evolution through the history. The observed cosmological data has no intrinsic meaning, unless there is an interpretation with a viable model that could generate knowledge. An attempt to predict the behavior of Hubble
parameter $H(z)$ as a function of redshift $z$ is a typical model. Different models may have different set of parameters and are not required to be physically motivated when defining a functional form for the Hubble parameter $H(z)$. The primary objective of interpretation of data involves estimation and optimization of various parameters appearing in the model under study. Often, we propose solutions with models of little or no cosmological or physical significance. Such attempts make model-independent predictions and are targeted to construct functions which provide better fit for the data. On the contrary, the model-dependent solutions are designed specifically for addressing various problems and anomalies which arise in theoretical and observational considerations. Since no model gives a complete understanding of the dynamics of the universe, with this study we intend to design an algorithm that can discriminate various models for pros and cons in an efficient way. This is accomplished by constraining the values of the parameters like Hubble constant $H_0=100h km s^{-1} Mpc^{-1}$, matter densities $\Omega_bh^2$ and $\Omega_ch^2$. Among these Hubble constant $(H_0)$ is very important in describing the current nature of the Universe since it represents the current expansion rate.

The discovery of accelerated expansion \cite{Perlmutter1999} along with the other observations necessitated the development of dark energy models to explain the universe's evolution as well as its origin. Among these various propositions one profound model which is now adopted as the concordance model is the $\Lambda CDM$ model. The model assumes a universe which is dominated by cold dark matter together with cosmological constant. And the Friedmann equation derived from the Einstein equation take the form,

\begin{align}
    \left(\frac{H}{H_0}\right)^2 & = \Omega_{m0}(z+1)^3 + \Omega_{r0}(z+1)^4 + \Omega_{\Lambda0}
\end{align}

with the condition,

\begin{align}
    1 & = \Omega_{m0} + \Omega_{r0} + \Omega_{\Lambda0}
\end{align}

\subsection{Type Ia Supernova observations and inferences}
\label{subsection:TyepIasupernova}

The SNe Ia observations are important probes in the precision cosmology as they trace out the expansion history of the universe with greater accuracy. What makes Type Ia supernovae good candidates for such experiments is the uniformity in their light curve. Although these objects are not standard candles, they are standardizable\cite{Wright2018}. In this study, we analyze the luminosity-distance relation dataset from the 1048 SNe spanning in a range of $0.01 < z < 2.3$ compiled by Scolnic et al.\cite{scolnic_complete_2018} also called as the ``Pantheon Sample``. Even though Type Ia supernova are considered  as standard candles, there are many inherent discrepancies.  Earlier discoveries suggest a substantial relationship between the supernova luminosity and the host galaxy. Likewise, there exist a similar relationship between the color and luminosity due to the influence of the surrounding medium, along with astrophysical reasons.  For each supernova, the light curve is fitted with three different parameters. The overall normalization $x_0$ which gives the time dependent spectral energy distribution of the supernova, the deviation from the average light curve $x_1$ and the deviation from average supernova B-V color. Consequently, the corrected distance modulus formula is given by,

\begin{align}
    \mu_B & = m^{\max}_{B}+ \alpha x_1 - \beta c \nonumber                    \\
          & + \delta P(m^{\text{true}}_{*} < m ^{\text{threshold}}_{*}) - M_B
\end{align}

where $ m^{\max}_{B}$ is the integrated B-band flux at maximum light, $M_B$ is the absolute B-band magnitude of a Type Ia supernova with $ x_1$ = 0, c = 0, and $P(m^{\text{true}}_{*} < m ^{\text{threshold}}_{*})$=0. Also, $m^{\text{threshold}}_{*} = 10^{10} M_{\odot} $ is the threshold host-galaxy mass used for the mass-dependent correction, and P is a probability function assigning a probability that the true mass, $m^{\text{true}}_{*}$, is less than the threshold value, when an actual mass measurement $m^{\text{obs}}_{*}$ is made\cite{amanullah_spectra_2010}.While fitting the light curve, the parameters  $\alpha,\beta, \delta$  and $M_B$  are called  “nuisance” parameters since they are dependent on the cosmological model assumed.

\subsection{CMB: experiments and inference}
The COBE experiment\cite{Smoot1992} was the first to provide observational confirmation for CMB and inflation, the CMB temperature curve was exactly that of a black body at $T= 2.7 K$ and an almost scale invariant power spectrum. Since then  the era of precision cosmology has brought upon tighter  constraints and higher understanding of the inflationary phase of universe.

The first detection of polarization of CMB was the result of observations from DASI telescope\cite{Halverson1998}. The E-mode CMB polarization signal was detected with $4.9 \sigma$ confidence level and the B-mode polarization  was given an upper limit of 0.59 at $95 \%$ confidence level. The next generation of experiments like CBI, WMAP  gave the finer limits for CMB temperature anisotropies. The measurements of relative heights of the acoustic peaks gives  the relative density of protons and dark matter and the angular power spectrum provided the temperature and polarization anisotropies  and also curvature of the universe.

The latest in these missions is the Planck mission\cite{Tauber2004}. The Planck results have fine tuned the cosmological parameter values with some speculation on observed anomalies like Cold spot and the Axis of evil. The repeated Planck mission has now showed that these anomalies have a very small significance level\cite{PlancK_2020}. Hence we have the parameter values as $H_0= 67.74$, $\Omega_{\Lambda}=0.69$, $\Omega_b h^2=0.04$, $\Omega_c h^2=0.25$, $\Omega_m= 0.3$ and $n_s=0.96$ which supports that $\Lambda$CDM model of the universe.

\section{Datasets}\label{section:datsets}
\subsection{Pantheon}
The Pantheon sample\cite{scolnic_complete_2018, koussour2022dynamical} is a compilation of 1048 SNe Ia data points\footnote{\url{https://archive.stsci.edu/prepds/ps1cosmo/scolnic_datatable.html}} from previous released datasets. In fact, it is the first sample containing $\sim O(10^3
    )$ SNe Ia. Hence, it is a very interesting pursuit to test cosmic anisotropy and other features of the universe with Pantheon compilation. The redshift range of these 1048 SNe Ia is given by,
$0.01012 \leq z \leq2.26$.
The theoretical distance modulus predicted by the cosmological model is defined by,
\begin{align}
    \mu_{\text{mod}}=5\log d_L+25
\end{align} where  $d_L=\frac{v_c}{H_o}D_L$ is the luminosity distance, $v_c$ is the speed of light and $H_o$ is the Hubble constant.
The luminosity distance is given by

\begin{align}
     & D_L(z_{cmb}, z_{hel})  =  (1+z_{hel})r(z_{cmb})                         \\
     & r(z_{cmb}) = cH^{-1}_0 \int^z_0 \frac{dz'}{E(z')}                       \\
     & E(z) = \Omega_r (1+z)^4 + \Omega_m(1+z)^3 + \Omega_{de}(1+z)^{3(1 + w)}
\end{align}
where $\Omega_r$, $\Omega_m$ and $\Omega_{de}$ are the radiation, matter and dark energy density parameters. Also,

\begin{align}
    \Omega_r = \Omega_m/(1+z_{eq})
\end{align}

$z_{\text{eq}} = 2.5 \times 10^4 \Omega_{m}h^2 (T_{\text{cmb}}/2.7K)^{-4}$, $T_{\text{cmb}} =2.7255K$ and $h$ is the reduced hubble constant\cite{Eisenstein1998}. The redshift $z$ is taken in two different frame of references, the heliocentric frame $z_{\text{hel}}$ and the CMB frame of reference $z_{\text{cmb}}$. The Chi-square estimation with the covariance matrix, $Cov \lbrace \mu (z_i), \mu (z_j)\rbrace$ is given as,

\begin{align}
    \chi^2 & =  \sum_{ij}\Delta \mu (z_i) Cov^{-1} \left\lbrace \mu (z_i), \mu (z_j) \right\rbrace \Delta \mu (z_j)
\end{align}

where,
\begin{align}
    \Delta \mu (z_i) = \mu(z_i)^{\text{obs}} - \mu(z_i|\theta)^{\text{theory}}
\end{align}
and $ \theta$ corresponds to the cosmological parameter to be estimated.

\subsection{Observational Hubble parameter Data (OHD)}
OHD\cite{Ma2011, pacif2021late} is widely used in model testing as well as in the analysis of different cosmological datasets as a tool for constraining cosmological parameters. They are not affected by the discrepancies of the model since the data set is obtained from model-independent direct observations. The distinctiveness of OHD is in its ability to compute H(z) without any integral terms occurring in the relation to be solved, hence rendering the expansion history directly via the observable. But this particular dataset unlike other cosmological observations is not homogeneous, instead, it comprises of three different classes of observations.

\subsubsection{Cosmic Chronometer (standard clock)}
In differential age method\cite{Moresco2012}, the difference in age is compared with the difference in red shift of any two galaxies that are formed at the same time as (dz/dt). Here the ages of passively-evolving galaxies are taken as the chronometers which is determined by spectroscopic dating. And the H(z) is computed directly from the differential age as, $H(z) = - \frac{1}{1+z} \frac{dz}{dt}$.

\subsubsection{Radial BAO size method (standard ruler)}
The detection of Baryon Acoustic peak in the galaxy power spectrum\cite{eisenstein2005detection,sanchez2006cosmological} opened a novel method to determine cosmological parameters which are highly affirmative to concordance model and the interpretations of CMB data. The Baryon Acoustic Oscillation (BAO) from the recombination epoch and its imprints on the peak along the radial direction of galaxy power spectrum is found to be significantly consistent as reported by  Gaztañaga et al.\cite{gaztanaga2009clustering}.  In this method, the BAO peak position is used as a standard ruler in the radial direction which enables us to get a direct measurement of the Hubble constant. The Hubble parameter $H(z)$ as a function of redshift can be directly determined using the radial measurements as, $H(z)_{true} = \frac{r_{BAO}}{r_{WMAP}} H(z)_{ref}$ with $H(z)_{ref} = H_0 \sqrt{0.25(1+z)^3 + 0.75}$.

\subsubsection{Gravitational Waves (standard siren)}
The success of LIGO and Virgo in analyzing the GW has opened a new method for obtaining the model-independent estimation of the Hubble parameter. The method named as ``Standard sirens method``\cite{Holz2005} employs analysis of gravitational waves from the neutron star (NS) binary systems/ neutron star, black hole mergers with the help of theoretical templates. Unlike the black holes, the NS binary system emanates electromagnetic radiation which is easily probed, allowing determination of the distribution of the system to be possible.

\subsection{Planck}

The products published by Planck mission project \cite{PlancK_2020} is a collection of CMB angular power spectra and likelihood functions. The CMB spectra is considered to be statistically isotropic which is supposed to be described by a multivariate gaussian. Also, the CMB angular power spectrum is assumed to be capable of revealing the fundamental parameters as well as the associated cosmological model that governs the dynamics of the primordial Universe. The Planck project provides a set of powerspectra corresponding to Temperature and Polarization. However, in our work we are analyzing the TT power spectrum only. The power spectrum ranges over multipoles $l=2-2508$. The process of estimating the power spectrum is strongly reliant on various models involving instrumental effects and other noises, hence a method involving pseudo-power spectrum is employed. The estimation comprises of two different methods and the components are separated for low-$l$ ($2<l<30$) and high-$l$ ($30<l<2500$) regions. The approximation for likelihood of these two, Commander and Plik respectively, are then combined to form the $TT$ spectrum of Planck mission. The data\footnote{\url{http://pla.esac.esa.int/pla/aio/product-action?COSMOLOGY.FILE_ID=COM_Likelihood_Data-baseline_R3.00.tar.gz}} and the likelihood functions, Planck likelihood code\footnote{\url{http://pla.esac.esa.int/pla/aio/product-action?COSMOLOGY.FILE_ID=COM_Likelihood_Code-v3.0_R3.10.tar.gz}}, are obtained from the Planck Legacy Archive.

\section{Methodology}\label{section:methodology}
When presented with a myriad of cosmological models, the essence of physics becomes computational validation of these models with observations. The methodology involved in inferring  cosmological parameters from large data compilations  with unprecedented uncertainties demands usage of  statistical methods. The major obstacle in designing a methodology for a cosmological problem is that the sample, the universe, we are working with is only one, hence a classical frequentist approach is no longer valid. A popular alternative approach is the Markov Chain Monte Carlo (MCMC), a technique based on the bayesian inference. In this section the strength and the weakness of the MCMC approach is reviewed and an alternative strategy as a suppliment to MCMC is proposed.

\subsection{MCMC based approach}
The objective of Bayesian inference is to estimate a posterior distribution, it mainly concerns with the observed data while frequentist approach deals with probability of data to be observed. The Bayes theorem is given as,

\begin{equation}
    P(\Theta|D, I) = \frac{P(D|\Theta, I)P(\Theta|I)}{P(D|I)}
\end{equation}

\begin{center}
    \resizebox{0.3 \linewidth}{!} {%
        \begin{tabular}{ll}
            $P(\Theta|D, I)$      & Posterior probability \\
            $P(D|\Theta, I)$      & Probability of data   \\
            $P(\Theta|I), P(D|I)$ & Prior probabilities
        \end{tabular}
    }
\end{center}

Markov Chain Monte Carlo exploits the bayesian inference to generate samples from the target distribution by  a stochastic process. The markov process in MCMC is a stochastic random walk method in which each sample depends only on the immediate previous sample.  The sampling is performed with the Metropolis-Hasting algorithm from a proposal distribution for the parameters. In MCMC, a parameter is a random variable with a probability distribution and the goal of the algorithm is to completely trace out the probability density of the distribution defining the variable. The transition kernel determines how the parameters move randomly to a new position in space. Let $Q$ be a distribution on $y$ given $x$. The transition kernel $Q(y|x)$ lets you move from $x$ to $y$. In most cases $Q$ is taken as a continuous distribution.

\begin{align}
    \int Q(y|x) dy & =1 \quad \forall x
\end{align}

The probability distribution is given by,
\begin{align}
    Q(y|x) & = \frac{1}{\sqrt{2 \pi}} \exp{{\left[-0.5(y-x)^2\right]}}
\end{align}

\subsubsection{MH Algorithm}
\label{section:MH}
\begin{itemize}
    \item Initialize $X_1 = x_1$
    \item for $t = 1,2, \ldots$
          \begin{itemize}
              \item Sample $y$ from $Q(y|x)$. Obtained $y$ is the proposed value for $x_{t+1}$ ($x_{t+1}=y$)
              \item Compute acceptance ratio, \\ $\alpha = min\left(1,\frac{ \Pi(y) Q(x_t|y)}{\Pi(x_t)Q(y|x_t)}\right)$
              \item $\alpha$ is often called the acceptance probability, which determines whether to
                    \begin{itemize}
                        \item accept $x_{t+1} = y$
                        \item or $x_{t+1} = x_t$
                    \end{itemize}
          \end{itemize}
\end{itemize}

\subsection{Genetic Algorithm}
Genetic algorithm (GA)\cite{Mirjalili2018} is part of evolutionary computation theories. It is used as  optimization algorithm, by imitating evolutionary features like mutations, Gene cross-over, natural selection etc. Like evolution, these processes are completely random but allows us to control the level of randomization and level of mixing. Genetic algorithms can work with all kinds of data, simultaneously test multiple points from the solution space, optimize with either discrete or continuous variables and provide multiple optimum solutions.

Genetic algorithm was propounded in 1970's and 1980's as an alternative for traditional optimization techniques and was developed as a  meta-heuristic technique to solve hybrid computational challenges.  John Holland\cite{Holland1992} had initially developed genetic algorithm based on Darwin's evolutionary theory which was further expanded by 1992.

GA are recognized to be part of statistical investigation processes which search for optimal solutions and also avoid common pitfalls. GA has many advantages over traditional methods. GA progresses from a pool of possible optimal solutions instead of moving from one possible solution to next thus  reducing the chances of converging to a local minima. The Optimization search only depends on numerical fitness values which are defined according to each problem hence making it easier to search through noisy, non linear and chaotic data. These algorithms can be fine tuned to get either accurate or more efficient results. As they don't require any extra information other than fitness functions  of the problem, this gives them an upper hand to other optimization techniques that cannot handle non linearity, lack of continuity or derivatives inherent in the problem.

\subsection{Novel hybrid Approach}
We are using modified meta-heuristic algorithm with Metropolis Acceptance Criterion for optimal parameter estimation. The algorithm is mostly based on the ideas of Genetic algorithm (GA) and Monte Carlo Markov Chain algorithm (MCMC). GA uses multiple intermixing chains in single run which results in improved stochasticity of the parameters being drawn. The Metropolis-Hasting criterion adapted from MCMC technique ensures inhibition of premature convergence to local minima. Hence an algorithm designed and developed with these principles would converge with less CPU cycles spent without sacrificing the accuracy of parameter convergence.

Also the requirement of huge computing power and time makes many stochastic algorithms less efficient with regard to computational performance. On contrary the dynamically evolving nature of evolutionary algorithms dominate such algorithms with less resource spent in converging to the solution with same precision. But the counter intuitive free evolution could possibly lead to unacceptable solutions often making it less accurate in practical applications. The novel design we propose here follows evolutionary design while keeping the balance between accuracy and precision by incorporating the famous MH criterion which is the central idea that makes MCMC reliable. Hence a novel Evolutionary Optimization of Cosmological Parameters using Metropolis Acceptance Criterion is developed.

Here we are presenting a case study to demonstrate the efficiency of our algorithm by estimating constraints for parameters of cosmological importance from the supernova data sets and the estimation cosmological parameters from CMB data provided by Planck mission. The Type Ia supernovae (SNe Ia) serves a great tool in probing and measuring dark energy and various parameters determining the course of evolution of universe. The high precision light curve data of SNe Ia enables an accurate construction of  magnitude-redshift relation over a greater range of redshift. Although recent studies were able to reduce statistical error in the datasets, the survey dependent systematic errors are unable to correct and are considered as nuisance parameters. As the test cases for demonstrating the efficacy of the algorithm developed, we will be using the data from 1048 SNe of the Pantheon compilation dataset, 52 $H(z)$ data and CMB data from Planck Mission.

\subsubsection{Algorithm}
\begin{itemize}
    \item Let the dataset be fitted with the parameters $\theta_1$ to $\theta_m$ where m gives the number of parameters of the model.
    \item An ensemble of  the set of possible combination of  parameters\footnote{One complete set of parameters represent a viable solution of the model to the dataset} is randomly selected from the given prior and is denoted by $\mathcal{K}$.

          \begin{align}
              \mathcal{K}= \{\phi^1,\phi^2,\phi^3, \ldots, \phi^n\}
          \end{align}

          where each $\phi^i=\{\theta^i_1,\theta^i_2, \ldots, \theta^i_m\}.$ The size of the ensemble $`n$' can be set according to the  precision of results required. Each iteration of algorithm gives a new ensemble which is called as  the Solution Ensemble. The log-likelihood of each solution $\phi^i$ is computed.
    \item A subset of $\mathcal{K}$, Seeder set $\mathcal{S}$ of size $N_s$ is  chosen to generate the Offspring Ensemble $\mathcal{O}$.

          \begin{align}
              \mathcal{S} \subset \mathcal{K}
          \end{align}

    \item Formation of $\mathcal{S}$ = $\{\psi^1,\psi^2, \ldots, \psi^{N_s}\}$
          \begin{itemize}
              \item The number of possible candidates in $\mathcal{S}$, $N_s$ is computed from the Seeder fraction value  $S_f$, given by $N_s$=$S_f*n$
              \item The possible candidates for the Seeder set $\mathcal{S}$ are staged by using the Tournament Selection and the MH - algorithm.
              \item A pair of candidate solutions is randomly chosen and the log-likelihood values are compared. The candidate with the highest log-likelihood $\mathcal{L}_{best}$ is taken as the $\phi^{best}$.
              \item The next candidate is accepted or rejected with respect to the Acceptance ratio described in Section \ref{section:MH} given as

                    \begin{align}
                        \alpha = min\left(1,\frac{ \Pi(y) Q(x_t|y)}{\Pi(x_t)Q(y|x_t)}\right)
                    \end{align}

              \item A random number is generated from the uniform distribution $U(0,1)$,  if this number is less than $\alpha$.  Then the current candidate solution is taken as $\phi^{best}$ otherwise the previous solution is retained.
              \item The process is repeated until the required ensemble is selected.

                    \begin{align}
                        \mathcal{S} = \{\psi^* \in \mathcal{K}|\psi^* \text{satisfies MH Tournament}\}
                    \end{align}

          \end{itemize}
    \item Formation of Offspring Ensemble $\mathcal{O}$
          \begin{itemize}
              \item Further the candidate solutions in $\mathcal{S}$ undergo Recombination and Migration.
              \item The Offspring Ensemble \\ $\mathcal{O}=\{O^1,O^2, \ldots, O^{N_s}\}$. Each Offspring $O^i$ is formed from two candidate solutions in $N_s$ i.e. by taking $\phi^i$ and $\phi^{i+1}$.
              \item Recombination- A random number $\beta$ is generated from uniform distribution $U(1,m)$.
              \item The candidate solutions $\phi^i$ and $\phi^{i+1}$ are broken at $\beta$ for the parameter solution. The initial portion of $\phi^i$ is added to final portion of $\phi^{i+1}$ , forming the offspring $O_i$.
              \item Migration- A subset $\mathcal{M}$ of size $N_m$ from the Offspring ensemble $\mathcal{O}$  is chosen randomly, where $N_m=M_f*n$ and $M_f$, migration fraction.

                    \begin{align}
                        \mathcal{M}=\{\mu^1,\mu^2, \ldots, \mu^{N_m}\}
                    \end{align}

              \item A set of random numbers $j$'s are generated from an uniform distribution $U(0,m)$ and  a single parameter in the location $j$ from each solution $\mu^i$ in the set $\mathcal{M}$ is selected. A value corresponds to the parameter $\theta^i_j$ is drawn from a normal distribution $N(\theta_j^i,\sigma_j)$ with $\sigma_j$ as the migration width and is replaced with the original value.
              \item Elite Population- Further the best candidate from the ensemble population $\mathcal{K}$ is directly added to the now formed offspring generation to bring the number candidate solutions equal to $`n$'.
          \end{itemize}
    \item Hence an Offspring population $O$ similar to the initial Ensemble population $\mathcal{K}$ is formed and these steps are repeated by taking the Offspring population as the new Ensemble population.
\end{itemize}

\section{Results and Discussion}\label{section:results}

We analyzed different observations provided by the Pantheon project, Observational Hubble Data and Planck mission. The best-fit parameters ($H_0$, $\Omega_m$, $\Omega_bh^2$, $\Omega_ch^2$, $\tau$, $A_s$, $n_s$ and $y_{cal}$) for SNeIa, OHD and CMB data were estimated against the $\Lambda CDM$ model which are found to be lying well within the limits. An interesting feature that was revealed during the process was the significantly lower computational time and steps, measured in the number of iteration and CMB anistropy simulation code calls required compared to the former methods employed in this regime. The number of iterations and sampling required in the parameter estimation process can be inferred from the trace plots for the different cases studied in this work.

\subsection{Estimation of cosmological parameters from the SNe Ia data - Pantheon Compilation}

\begin{figure}
    \centering
    \begin{subfigure}[t]{0.9\textwidth}
        \centering
        \includegraphics[width=\textwidth]{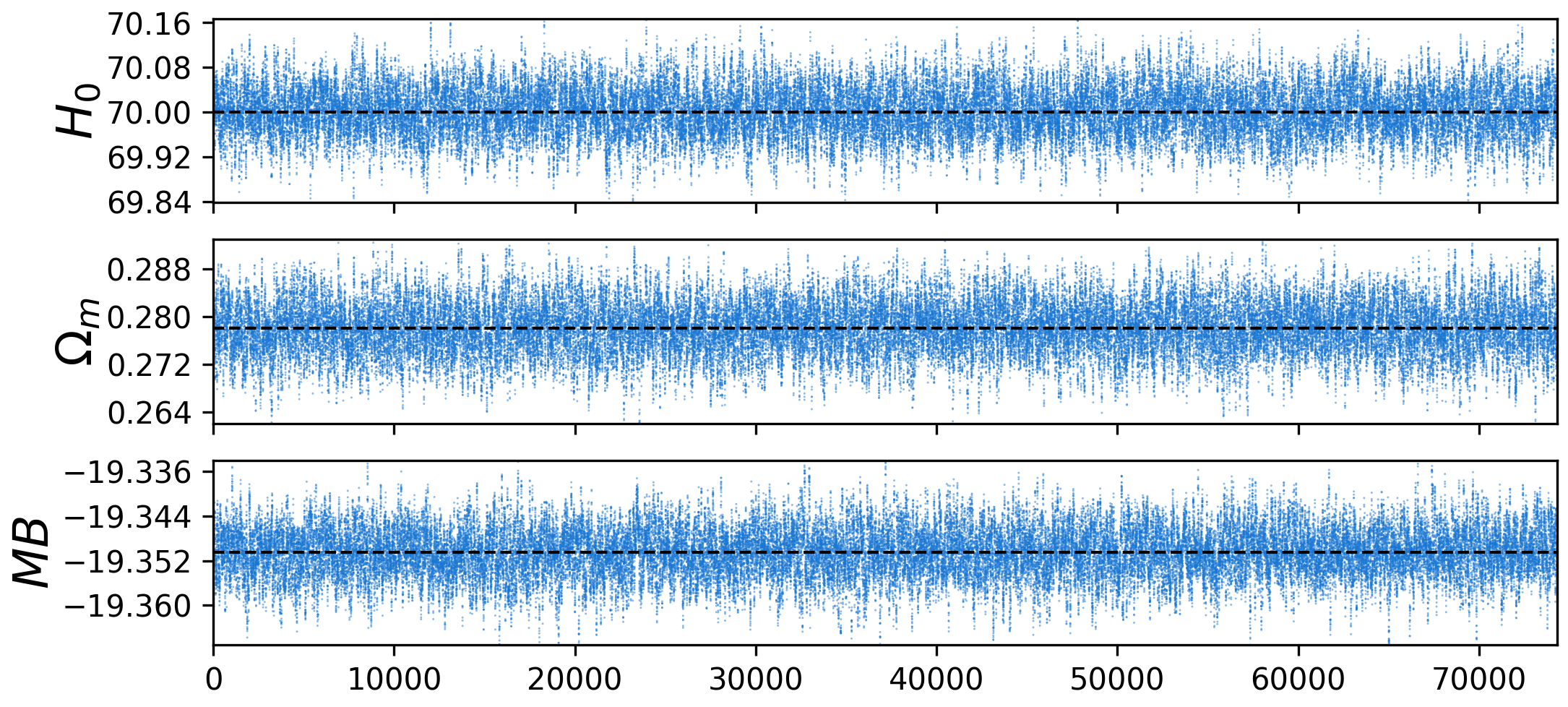}
        \caption{Trace plot of $H_0$ and $\Omega_m$ from the chain.}
    \end{subfigure}
    \hfill
    \begin{subfigure}[t]{0.45\textwidth}
        \centering
        \includegraphics[width=\textwidth]{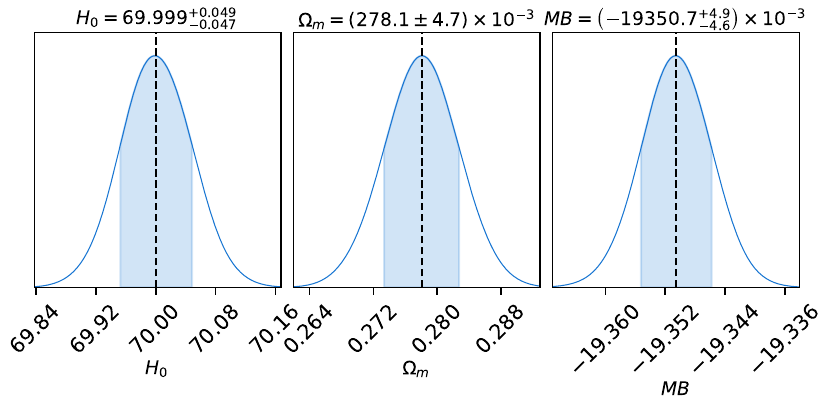}
        \caption{Marginal distributions of $H_0$ and $\Omega_m$ for Pantheon data.}
    \end{subfigure}
    \caption{Analysis for parameters of $\Lambda CDM$ model for Pantheon data from the samples obtained from the iterations of the algorithm.}
    \label{fig:pantheon}
\end{figure}

The $\Lambda CDM$ model is confronted with the pantheon dataset which consist of 1048 SNeIa observations which is shown in Fig.\ref{fig:pantheon}. The figure shows the trajectories of parameter values which clearly indicates the convergence of the estimation chain with the values $H_0=69.999^{+0.049}_{-0.047}$, $\Omega_m=0.2781 \pm 0.0047$ and the absolute magnitude value $MB=-19.3507^{+0.0049}_{-0.0046}$ \cite{scolnic_complete_2018}. The values obtained shows consistency with the previous results\cite{bouali2023model} ($H_0\approx70kms^{-1}Mpc^{-1}$ and $\Omega_m=0.284\pm0.012$) from Mehrabi\cite{Mehrabi2020}, ($H_0=70.0kms^{-1}Mpc^{-1}$ and $\Omega_m=0.25$) Thakur\cite{Thakur2021} and Scolnic \cite{scolnic_complete_2018}.

\subsection{Insights from Observational Hubble Parameter Data (OHD)}

\begin{figure}
    \centering
    \begin{subfigure}[t]{0.9\textwidth}
        \centering
        \includegraphics[width=\textwidth]{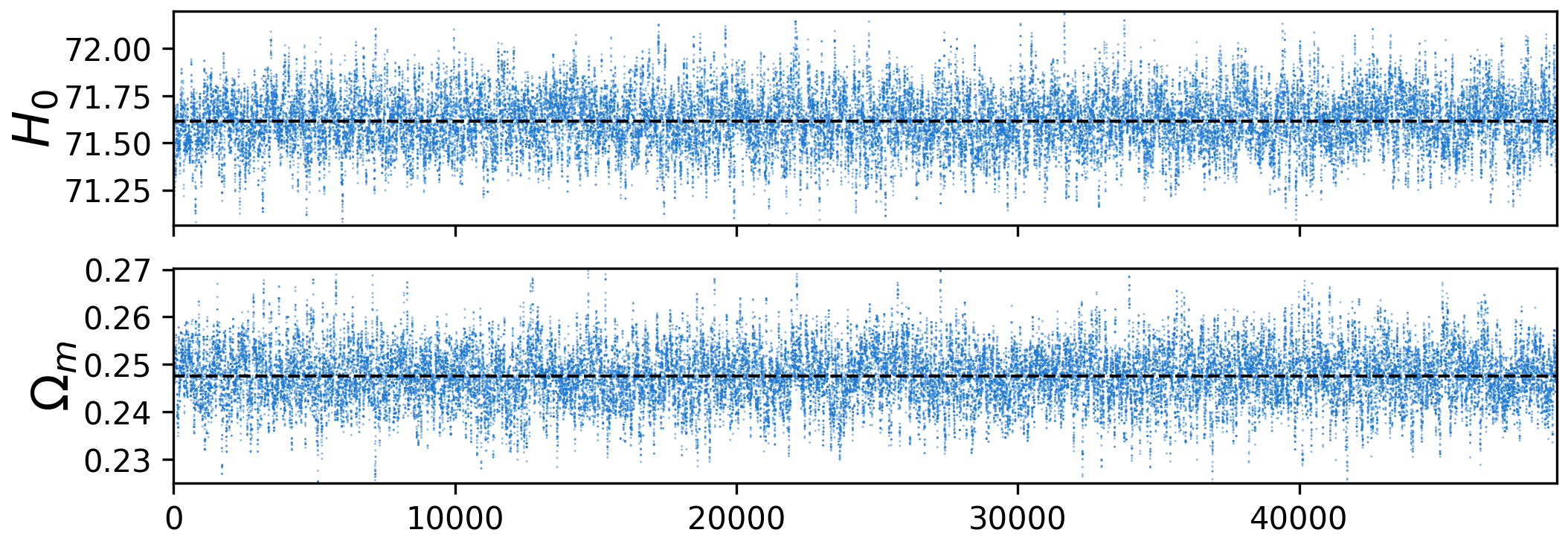}
        \caption{Trace plot of $H_0$ and $\Omega_m$ from the chain for OHD.}
    \end{subfigure}
    \hfill
    \begin{subfigure}[t]{0.45\textwidth}
        \centering
        \includegraphics[width=\textwidth]{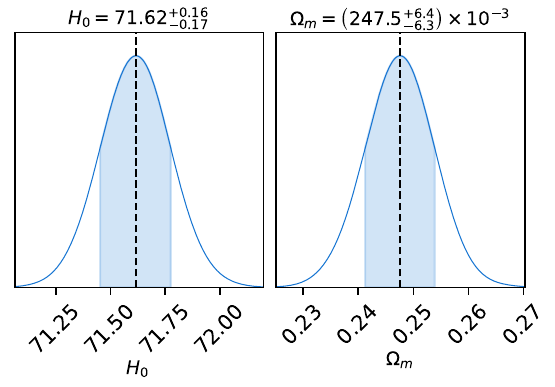}
        \caption{Marginal distributions of $H_0$ and $\Omega_m$ for $\Lambda CDM$.}
    \end{subfigure}
    \caption{Analysis for parameters of $\Lambda CDM$ model for Observational Hubble parameter data from the samples obtained from the evolutionary algorithm.}
    \label{fig:ohd_chain}
\end{figure}

The analysis of observational hubble parameter data with our evolutionary algorithm has obtained the parameters of the $\Lambda CDM$ model of the universe Fig.\ref{fig:ohd_chain}. The parameter values, $H_0=71.62^{+0.16}_{-0.17}$ and $\Omega_m=0.2475^{+0.0064}_{-0.0063}$, obtained exhibits consistency with other previous estimations ($H_0=72kms^{-1}Mpc^{-1}$ and $\Omega_m=0.25$) made on OHD\cite{zhang2010constraints,geng2018prospect,gangopadhyay2023generic}.

\subsection{Analysis of CMB data and parameter estimation with Planck}

\begin{figure}
    \centering
    \begin{subfigure}[t]{0.7\textwidth}
        \centering
        \includegraphics[width=\textwidth]{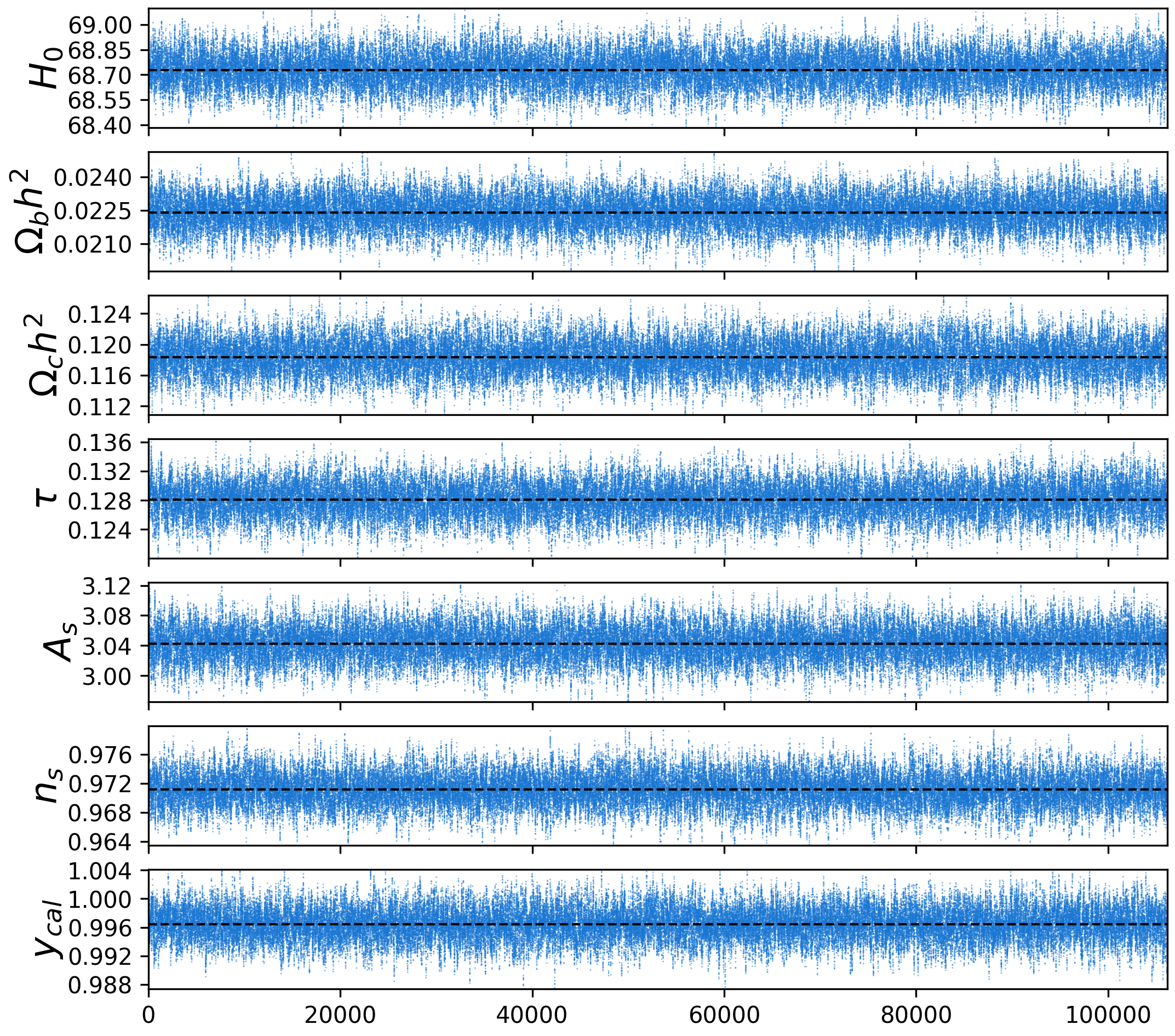}
        \caption{Trace plot of $H_0$, $\Omega_{b}h^2$, $\Omega_{c}h^2$, $\tau$, $\rm{ln}(10^{10} A_s)$, $n_s$ and $y_{cal}$ from the chains for Planck CMB data.}
    \end{subfigure}
    \hfill
    \begin{subfigure}[t]{0.9\textwidth}
        \centering
        \includegraphics[width=\textwidth]{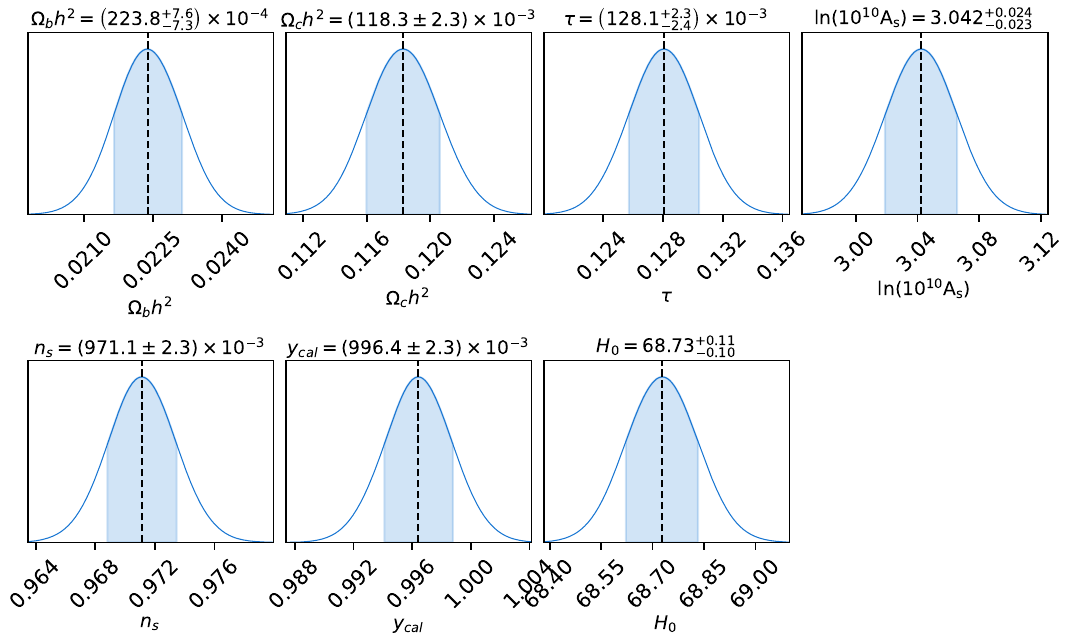}
        \caption{Marginal distributions of $H_0$, $\Omega_{b}h^2$, $\Omega_{c}h^2$, $\tau$, $\rm{ln}(10^{10} A_s)$, $n_s$ and $y_{cal}$ for $\Lambda CDM$.}
    \end{subfigure}
    \caption{Analysis for parameters of $\Lambda CDM$ model for Planck data from the chains obtained from the evolutionary algorithm.}
    \label{fig:plc_chain}
\end{figure}

We performed an exploration of the 6 parameter space of $\Lambda CDM$ model with Planck CMB data. Fig.\ref{fig:plc_chain} shows a certain convergence of the stochastic estimation of the parameters to the values $H_0=68.73^{+0.11}_{-0.10}$, $\Omega_{b}h^2=0.02238^{+0.00076}_{-0.00073}$, $\Omega_{c}h^2=0.1183 \pm 0.0023$, $\tau=0.1281 \pm 0.0023$, $\rm{ln}(10^{10} A_s)=3.042^{+0.024}_{-0.023}$, $n_s=0.9711 \pm 0.0023$ and $y_{cal}=0.9964 \pm 0.0023$. Comparison of the obtained values with those estimated by other authors\cite{PlancK_2020,DiValentino2016} indicates sufficient consistency and the computational efficiency of the method. In this excercise we used the publicly availabe data and likelihood code \textit{plik\_rd12\_HM\_v22\_TT} provided by Planck mission. The results obtained from the CosmoMC program and our method are compared in the Table.\ref{table:plc_results}.

\begin{table}
    \centering
    \begin{tabular}{lll}
        \hline
        \hline
                               & Evolutionary Algorithm          & CosmoMC                   \\
        \hline
        $H_0$                  & $68.73^{+0.11}_{-0.10}$         & $68.1\pm 1.2$             \\ \\
        $\Omega_bh^2$          & $0.02238^{+0.00076}_{-0.00073}$ & $0.02243\pm 0.00028$      \\ \\
        $\Omega_ch^2$          & $0.1183 \pm 0.0023$             & $0.1182\pm 0.0027$        \\ \\
        $\tau$                 & $0.1281 \pm 0.0023$             & $0.129^{+0.037}_{-0.033}$ \\ \\
        $\rm{ln}(10^{10} A_s)$ & $3.042^{+0.024}_{-0.023}$       & $3.188^{+0.070}_{-0.061}$ \\ \\
        $n_s$                  & $0.9711 \pm 0.0023$             & $0.9706\pm 0.0082$        \\ \\
        $y_{cal}$              & $0.9964 \pm 0.0023$             & $1.0002\pm 0.0025$        \\
        \hline
    \end{tabular}
    \caption{Comparison of cosmological parameters and their uncertainties obtained from evolutionary algorithm and CosmoMC for $\Lambda CDM$ model}
    \label{table:plc_results}
\end{table}

\newpage

\section{Conclusion}\label{section:conclusion}
We have introduced a new approach to cosmological parameter estimation and performed the computational analysis with Pantheon, OHD, Planck datasets. The algorithm we designed is implemented with a combination of Python and C++ languages. The computational efficiency and accuracy in the determination of various parameters are evident from the results described in Section \ref{section:results}. An interesting result revealed in our analysis is the discrepancy in the value of $H_0$, which is inclined to $67$ for the Planck CMB dataset and towards $70$ for the other two datasets, indicating Hubble tension. Unlike MCMC the hyper parameters of our algorithm which is derived from the design principles of Genetic Algorithm are easily tuned allowing us to efficiently converge to global or local extremum. This allows more freedom and control over the execution flow which is one of the advantage of using this algorithm. Hence this newly proposed evolutionary algorithm can be implemented independently or as a supplement to MCMC implementations. Furthermore, this also allows having a larger search space for the parameters that are to be estimated, which avoids stringent constraints over the parameters that are under scrutiny.

\section*{Acknowledgements}
We thank Titus K. Mathew and Charles Jose for the fruitful discussions and suggestions. This work utilised the research facility funded by  UGC-BSR start-up grant of CJ (No. F./30-463/2019(BSR)). SPS acknowledges the financial support by UGC, RGNF (No.: F1-17.1/2016-17/RGNF-2015-17-SC-KER-4135). MJ acknowledges the Associateship of IUCAA.


\begin{thebibliography}{00}
    \bibitem{Hinshaw2013}
    G.~Hinshaw, D.~Larson, E.~Komatsu, D.N.~Spergel, C.L.~Bennett, J.~Dunkley
    et~al., \emph{{Nine}-{year}{Wilkinson} {microwave} {anisotropy}
    {probe}({WMAP}) {observations}: {Cosmological} {parameter}{results}},
    \href{https://doi.org/10.1088/0067-0049/208/2/19}{\emph{The Astrophysical
            Journal Supplement Series} {\bfseries 208} (2013) 19}.

    \bibitem{Tauber2004}
    J.~Tauber, \emph{The planck mission},
    \href{https://doi.org/10.1016/j.asr.2003.05.025}{\emph{Advances in Space
            Research} {\bfseries 34} (2004) 491}.

    \bibitem{crill2003boomerang}
    B.~Crill, P.A.~Ade, D.~Artusa, R.~Bhatia, J.~Bock, A.~Boscaleri et~al.,
    \emph{Boomerang: A balloon-borne millimeter-wave telescope and total power
        receiver for mapping anisotropy in the cosmic microwave background},
    {\emph{The Astrophysical Journal Supplement Series} {\bfseries 148} (2003)
            527}.

    \bibitem{rabii2006maxima}
    B.~Rabii, C.D.~Winant, J.~Collins, A.~Lee, P.~Richards, M.~Abroe et~al.,
    \emph{Maxima: A balloon-borne cosmic microwave background anisotropy
        experiment}, {\emph{Review of Scientific Instruments} {\bfseries 77} (2006)
            071101}.

    \bibitem{aiola2020atacama}
    S.~Aiola, E.~Calabrese, L.~Maurin, S.~Naess, B.L.~Schmitt, M.H.~Abitbol et~al.,
    \emph{The atacama cosmology telescope: Dr4 maps and cosmological parameters},
    {\emph{Journal of Cosmology and Astroparticle Physics} {\bfseries 2020}
            (2020) 047}.

    \bibitem{chown2018maps}
    R.~Chown, Y.~Omori, K.~Aylor, B.~Benson, L.~Bleem, J.~Carlstrom et~al.,
    \emph{Maps of the southern millimeter-wave sky from combined 2500 deg2 spt-sz
        and planck temperature data}, {\emph{The Astrophysical Journal Supplement
                Series} {\bfseries 239} (2018) 10}.

    \bibitem{York2000}
    D.G.~York, J.~Adelman, J.~John E.~Anderson, S.F.~Anderson, J.~Annis,
    N.A.~Bahcall et~al., \emph{The sloan digital sky survey: Technical summary},
    \href{https://doi.org/10.1086/301513}{\emph{The Astronomical Journal}
        {\bfseries 120} (2000) 1579}.

    \bibitem{brooks2011handbook}
    S.~Brooks, A.~Gelman, G.~Jones and X.~Meng, \emph{Handbook of Markov Chain
        Monte Carlo}, Chapman \& Hall/CRC Handbooks of Modern Statistical Methods,
    CRC Press (2011).

    \bibitem{Lewis_2002}
    A.~Lewis and S.~Bridle, \emph{Cosmological parameters from cmb and other data:
        A monte carlo approach},
    \href{https://doi.org/10.1103/physrevd.66.103511}{\emph{Physical Review D}
        {\bfseries 66} (2002) }.

    \bibitem{Audren:2012wb}
    B.~Audren, J.~Lesgourgues, K.~Benabed and S.~Prunet, \emph{{Conservative
                Constraints on Early Cosmology: an illustration of the Monte Python
                cosmological parameter inference code}},
    \href{https://doi.org/10.1088/1475-7516/2013/02/001}{\emph{JCAP} {\bfseries
            1302} (2013) 001} [\href{https://arxiv.org/abs/1210.7183}{{\ttfamily
                    1210.7183}}].

    \bibitem{zuntz2015cosmosis}
    J.~Zuntz, M.~Paterno, E.~Jennings, D.~Rudd, A.~Manzotti, S.~Dodelson et~al.,
    \emph{Cosmosis: Modular cosmological parameter estimation}, {\emph{Astronomy
                and Computing} {\bfseries 12} (2015) 45}.

    \bibitem{torrado2021cobaya}
    J.~Torrado and A.~Lewis, \emph{Cobaya: Code for bayesian analysis of
        hierarchical physical models}, {\emph{Journal of Cosmology and Astroparticle
                Physics} {\bfseries 2021} (2021) 057}.

    \bibitem{metropolis}
    S.~Chib and E.~Greenberg, \emph{Understanding the metropolis-hastings
        algorithm}, \href{https://doi.org/10.1080/00031305.1995.10476177}{\emph{The
            American Statistician} {\bfseries 49} (1995) 327}.

    \bibitem{geyer_practical_1992}
    C.J.~Geyer, \emph{Practical {Markov} {Chain} {Monte} {Carlo}},
    {\emph{Statistical Science} {\bfseries 7} (1992) 473}.

    \bibitem{raftery_practical_1992}
    A.E.~Raftery and S.M.~Lewis, \emph{[{Practical} {Markov} {Chain} {Monte}
                {Carlo}]: {Comment}: {One} {Long} {Run} with {Diagnostics}: {Implementation}
    {Strategies} for {Markov} {Chain} {Monte} {Carlo}}, {\emph{Statistical
                Science} {\bfseries 7} (1992) 493}.

    \bibitem{simon2013evolutionary}
    D.~Simon, \emph{Evolutionary Optimization Algorithms}, Wiley (2013).

    \bibitem{Perlmutter1999}
    S.~Perlmutter, M.S.~Turner and M.~White, \emph{Constraining dark energy with
        type ia supernovae and large-scale structure},
    \href{https://doi.org/10.1103/physrevlett.83.670}{\emph{Physical Review
            Letters} {\bfseries 83} (1999) 670}.

    \bibitem{Wright2018}
    B.S.~Wright and B.~Li, \emph{Type ia supernovae, standardizable candles, and
        gravity}, \href{https://doi.org/10.1103/physrevd.97.083505}{\emph{Physical
            Review D} {\bfseries 97} (2018) }.

    \bibitem{scolnic_complete_2018}
    D.M.~Scolnic, D.O.~Jones, A.~Rest, Y.C.~Pan, R.~Chornock, R.J.~Foley et~al.,
    \emph{The complete light-curve sample of spectroscopically confirmed {SNe}
        {Ia} from {Pan}-{STARRS1} and cosmological constraints from the combined
        {Pantheon} sample}, \href{https://doi.org/10.3847/1538-4357/aab9bb}{\emph{The
            Astrophysical Journal} {\bfseries 859} (2018) 101}.

    \bibitem{amanullah_spectra_2010}
    R.~Amanullah, C.~Lidman, D.~Rubin, G.~Aldering, P.~Astier, K.~Barbary et~al.,
    \emph{Spectra and hubble space telescope light curves of six type {Ia}
    supernovae at 0.511 {\textless}z{\textless} 1.12 and the union2 compilation},
    \href{https://doi.org/10.1088/0004-637x/716/1/712}{\emph{The Astrophysical
            Journal} {\bfseries 716} (2010) 712}.

    \bibitem{Smoot1992}
    G.F.~Smoot, C.L.~Bennett, A.~Kogut, E.L.~Wright, J.~Aymon, N.W.~Boggess et~al.,
    \emph{Structure in the {COBE} differential microwave radiometer first-year
        maps}, \href{https://doi.org/10.1086/186504}{\emph{The Astrophysical Journal}
        {\bfseries 396} (1992) L1}.

    \bibitem{Halverson1998}
    N.~Halverson, J.E.~Carlstrom, M.~Dragovan, W.L.~Holzapfel and J.~Kovac,
    \emph{Dasi: a degree angular scale interferometer for imaging anisotropy in
        the cosmic microwave background},  in \emph{Advanced Technology {MMW}, Radio,
        and Terahertz Telescopes}, T.G.~Phillips, ed., {SPIE}, July, 1998,
    \href{https://doi.org/10.1117/12.317374}{DOI}.

    \bibitem{PlancK_2020}
    N.~Aghanim, Y.~Akrami, M.~Ashdown, J.~Aumont, C.~Baccigalupi, M.~Ballardini
    et~al., \emph{Planck 2018 results},
    \href{https://doi.org/10.1051/0004-6361/201936386}{\emph{Astronomy \&
            Astrophysics} {\bfseries 641} (2020) A5}.

    \bibitem{Eisenstein1998}
    D.J.~Eisenstein and W.~Hu, \emph{Baryonic features in the matter transfer
        function}, \href{https://doi.org/10.1086/305424}{\emph{The Astrophysical
            Journal} {\bfseries 496} (1998) 605}.

    \bibitem{Ma2011}
    C.~Ma and T.-J.~Zhang, \emph{{Power} {of} {observational} {Hubble} {parameter}
            {data}: A {figure} {of} {merit} {exploration}},
    \href{https://doi.org/10.1088/0004-637x/730/2/74}{\emph{The Astrophysical
            Journal} {\bfseries 730} (2011) 74}.

    \bibitem{Moresco2012}
    M.~Moresco, A.~Cimatti, R.~Jimenez, L.~Pozzetti, G.~Zamorani, M.~Bolzonella
    et~al., \emph{Improved constraints on the expansion rate of the universe up
        toz$\sim$ 1.1 from the spectroscopic evolution of cosmic chronometers},
    \href{https://doi.org/10.1088/1475-7516/2012/08/006}{\emph{Journal of
            Cosmology and Astroparticle Physics} {\bfseries 2012} (2012) 006}.

    \bibitem{eisenstein2005detection}
    D.J.~Eisenstein, I.~Zehavi, D.W.~Hogg, R.~Scoccimarro, M.R.~Blanton,
    R.C.~Nichol et~al., \emph{Detection of the baryon acoustic peak in the
        large-scale correlation function of sdss luminous red galaxies}, {\emph{The
                Astrophysical Journal} {\bfseries 633} (2005) 560}.

    \bibitem{sanchez2006cosmological}
    A.G.~S{\'a}nchez, C.~Baugh, W.~Percival, J.~Peacock, N.~Padilla, S.~Cole
    et~al., \emph{Cosmological parameters from cosmic microwave background
        measurements and the final 2df galaxy redshift survey power spectrum},
    {\emph{Monthly Notices of the Royal Astronomical Society} {\bfseries 366}
            (2006) 189}.

    \bibitem{gaztanaga2009clustering}
    E.~Gaztanaga, A.~Cabr{\'e} and L.~Hui, \emph{Clustering of luminous red
        galaxies--iv. baryon acoustic peak in the line-of-sight direction and a
        direct measurement of h (z)}, {\emph{Monthly Notices of the Royal
                Astronomical Society} {\bfseries 399} (2009) 1663}.

    \bibitem{Holz2005}
    D.E.~Holz and S.A.~Hughes, \emph{Using gravitational-wave standard sirens},
    \href{https://doi.org/10.1086/431341}{\emph{The Astrophysical Journal}
        {\bfseries 629} (2005) 15}.

    \bibitem{Mirjalili2018}
    S.~Mirjalili, \emph{Genetic algorithm},  in \emph{Studies in Computational
        Intelligence}, pp.~43--55, Springer International Publishing (2018),
    \href{https://doi.org/10.1007/978-3-319-93025-1_4}{DOI}.

    \bibitem{Holland1992}
    J.H.~Holland, \emph{Genetic algorithms}, {\emph{Scientific American} {\bfseries
                267} (1992) 66}.

    \bibitem{Mehrabi2020}
    A.~Mehrabi and S.~Basilakos, \emph{Does {$\Lambda CDM$} really be in tension
        with the hubble diagram data?},
    \href{https://doi.org/10.1140/epjc/s10052-020-8221-2}{\emph{The European
            Physical Journal C} {\bfseries 80} (2020) }.

    \bibitem{zhang2010constraints}
    T.-J.~Zhang, C.~Ma and T.~Lan, \emph{Constraints on the dark side of the
        universe and observational hubble parameter data}, {\emph{Advances in
                Astronomy} {\bfseries 2010} (2010) }.

    \bibitem{geng2018prospect}
    J.-J.~Geng, R.-Y.~Guo, A.-Z.~Wang, J.-F.~Zhang and X.~Zhang, \emph{Prospect for
        cosmological parameter estimation using future hubble parameter
        measurements}, {\emph{Communications in Theoretical Physics} {\bfseries 70}
            (2018) 445}.

    \bibitem{DiValentino2016}
    E.D.~Valentino, A.~Melchiorri and J.~Silk, \emph{Reconciling planck with the
        local value of h 0 in extended parameter space},
    \href{https://doi.org/10.1016/j.physletb.2016.08.043}{\emph{Physics Letters
            B} {\bfseries 761} (2016) 242}.

    \bibitem{Hinton2016}
    S.R.~{Hinton}, \emph{{ChainConsumer}},
    \href{https://doi.org/10.21105/joss.00045}{\emph{The Journal of Open Source
            Software} {\bfseries 1} (2016) 00045}.

    \bibitem{Thakur2021}
    R.K.~Thakur, M.~Singh, S.~Gupta and R.~Nigam, \emph{Cosmological analysis using Panstarrs data: Hubble constant and direction dependence}, {\emph{Physics of the Dark Universe} {\bfseries 34} (2021) 100894}.
    
    \bibitem{pacif2021late}
Pacif, SKJ, Arora, Simran, Sahoo, PK.
\emph{Late-time acceleration with a scalar field source: Observational constraints and statefinder diagnostics}.
\emph{Physics of the Dark Universe} \textbf{32} (2021) 100804.

\bibitem{koussour2022dynamical}
Koussour, M, Pacif, SKJ, Bennai, M, Sahoo, PK.
\emph{Dynamical dark energy models from a new Hubble parameter in $f(Q)$ gravity}.
\emph{arXiv preprint arXiv:2208.04723} (2022).

\bibitem{gangopadhyay2023generic}
Gangopadhyay, Mayukh R, Pacif, Shibesh K Jas, Sami, Mohammad, Sharma, Mohit K.
\emph{Generic modification of gravity, late time acceleration and Hubble tension}.
\emph{Universe} \textbf{9}(2), 83 (2023).

\bibitem{bouali2023model}
Bouali, Amine, Chaudhary, Himanshu, Mehrotra, Amritansh, Pacif, SKJ.
\emph{Model-independent study for a quintessence model of dark energy: Analysis and Observational constraints}.
\emph{arXiv preprint arXiv:2304.02652} (2023).

\end{thebibliography}
\end{document}